\documentstyle[prl,epsfig,aps]{revtex}
\def \bi{\bibitem}

\def\d{{\rm d}}
 \def\(({\left(}
 \def\)){\right)}
\def\bi{\bibitem}

\def \a{\alpha}
\def \b{\beta}

\def \d{{\rm d}}

\def \beqna{\begin{eqnarray}}
\def \eeqna{\end{eqnarray}}
\def \beq{\begin{equation}}
\def \eeq{\end{equation}}

\def \a{\alpha}

\def \b{\beta}

\def \ab2{\alpha\beta^2}

 \newcommand \be {\begin{equation}}
\newcommand \bea {\begin{eqnarray} \nonumber }
\newcommand \ee {\end{equation}}
\newcommand \eea {\end{eqnarray}}

\input{epsf}
\begin{document}
%\twocolumn[\hsize\textwidth\columnwidth\hsize\csname@twocolumnfalse\endcsname
%\draft
\narrowtext 
\twocolumn
\title{Critical properties of a  three dimensional p-spin model}
\author{ Silvio Franz(*) and Giorgio Parisi(**) }
\address{
(*) Abdus Salam International Center for Theoretical Physics\\
Strada Costiera 11,
P.O. Box 563,
34100 Trieste (Italy)\\
(**) Universit\`a di Roma ``La Sapienza''\\
Piazzale A. Moro 2,
00185 Rome (Italy)\\
e-mail: {\it franz@ictp.trieste.it }\\
e-mail: {\it parisi@roma1.infn.it}
}
\date{January 1997}
\maketitle
%\quad PACS. 75.24 M-- Numerical simulation studies.\\
%\quad PACS. 75.50 L-- Spin glasses.
\begin{abstract}
In this paper we study the critical  properties of 
a finite dimensional generalization of the $p$-spin model.
We find evidence that in dimension three, contrary to its mean field limit, 
the glass transition is associated to a diverging susceptibility
(and correlation length). 
\end{abstract}
\section{Introduction}
Two different transition mechanisms are known in spin glass 
mean-field theory \cite{mpv,young}, according to the form  
of the random Hamiltonian. 
In some  models, 
like the infinite range Sherrington-Kirkpatrick (S-K) model for spin glasses,
 there is a second order glassy transition with 
diverging spin-glass susceptibility and continuous replica 
symmetry breaking. In other  models,
 whose 
prototype is the Random Energy Model,  the transition is {\sl first order}
with a Gibbs-Dimarzio like entropy crisis. Other examples of models with the
 second type of behaviour are spin models with p-spin interaction, with $p>2$,
both for Ising spins and for spherical spins.     

 Numerical 
simulations indicate that the first type of mechanism 
describes the ergodicity breaking transition of 
finite dimensional spin glasses \cite{MPR}.
 The second mechanism is more appropriate 
to describe the behavior of structural glasses\cite{kitiwo}. 
The passage from mean-field
to finite dimension is in both cases highly non trivial. Despite many 
progresses \cite{dedoko} the problem of including fluctuations in the 
description of the finite dimensional spin glass transition
is far from being achieved. For that reason the test of the mean-field 
picture has been left in the last 15 years to the numerical study
of the Edwards-Anderson model, which admits the SK as infinite dimensional 
limit. 

Strangely enough there are only few numerical studies of finite 
dimensional spin  models that could have a transition homologous to 
the mean-field discontinuous transition. Given the relevance of this 
transition to structural glasses, the 
study of such finite dimensional models is of primary importance.

Up recently,to our knowledge, the only studies appeared in the literature, 
are these references \cite{HR} and \cite{frarit}. In reference
\cite{frarit} it was proposed a generalization of the $p$-spin model
which presents a phenomenology reminiscent to that of structural glasses.
However, the difficulty to decide about the existence of a phase transition
and the presence of a spurious symmetry, makes necessary to resort to 
better conceived models without spurious effects.  

In this paper we introduce, and study numerically, a finite dimensional 
model with $N$ spin per sites and $p$-spin interactions, that for all
dimensionality tends to a mean-field behaviour  for $N\to \infty$.
As it happens in ordinary field-theory, this large $N$ approach 
is complementary to the high-dimensionality approach. We study this
model for $p=4$ in $D=3$. A complementary study of
the same model (still for $p=4$ and $D=3$)
in the low temperature regime can be found in ref.
\cite{BACAPA}. Some numerical simulation of the $p=3$ model for $D=4$ 
can be found in ref. \cite{SEI}.

\section{A finite dimensional $p$-spin model}

The long range p-spin model \cite{pspin}
is soluble and its Hamiltonian is given by
\be 
H=-\sum_{i_1<...<i_p}^{1,N} J_{i_1,...,i_p}S_{i_1}...S_{i_p},
\label{h1}
\ee
where the variables $J$ are random with zero average and variance $1/N^{(p-1)}$
(the same result is obtained for Gaussian distributed variables and 
in the case $J=\pm 1/N^{((p-1)/2)}$) and the spins are Ising variables 
(also the 
spherical case is soluble).

This Hamiltonian  
can be generalized in many ways in finite $D$. The way we follow in this 
paper is the following: we consider a $D$ dimensional square lattice 
with $N$ Ising spins $S_x^\a$ ($\a=1,...,N)$ in each site $x$ of the lattice. 
For any given couples of nearest neighbour sites 
there are
$(2N)!/p!(2N-p)!$ possible groups $g$ of p-spins. We consider 
the product of all the spins in each group, and we couple them with 
a random variable $J_g$. The resulting Hamiltonian,
with transparent notation, is
\be 
H=-
\sum_{<x,y>} 
\sum_{g\in (x,y)} J_g \prod_{\mu\in g}S_\mu,
\ee
where we have relabeled the spins. The $J_p$ are chosen 
independently from link to link and group to group and
are equal to $\pm 1$ with probability $1/2$. The mean field 
limit is recovered both for high dimension ($D\to \infty$)  and  
finite $N$ or for large $N$ and finite $D$. Indeed it is possible
to construct  a loop expansion for the development in powers of 
$1/N$ \cite{capara}. In this paper 
we present a numerical study of the model in $D=3$ for 
low values of $N$ above the transition point.

We have simulated  the physics of the model through 
Monte Carlo method with the Metropolis algorithm. 
We have studied in detail the cases $p=4$ and  $N=3$ and 4.
As we will see the results do not seem to depend qualitatively on $N$ 
in this range of $N$ (for $N=2$ and $p=4$ the model is isomorphic to the 
usual short range Edwards Anderson model for spin glasses).  

In this first study of the model we discuss mainly three issues:
\begin{itemize}
\item
The existence of a glass transition by means of the study of the 
thermodynamics of the model, through the behaviour of the energy and 
the entropy in simulations of ``cooling experiments''.

\item
The behaviour of the time dependent auto-correlation function
at equilibrium and the growing of the relaxation time 
as the glass transition is approached.

\item
The existence of a growing spin-glass correlation length.
\end{itemize}

\begin{figure}
 \epsfxsize=250pt
\begin{center}
\epsffile{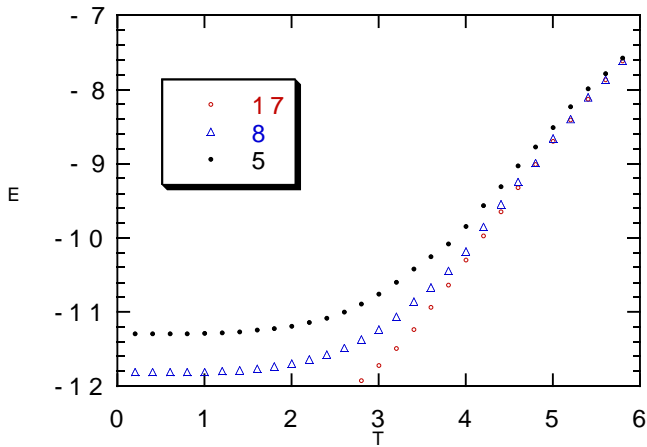} 
\end{center}
   \caption[0]{\protect\label{energie}  
The energy for $N=3$ and three different values of the cooling 
rate $\kappa=2^{-5},2^{-8},2^{-17}$. 
 } \end{figure}

\begin{figure}
 \epsfxsize=250pt
\begin{center}
\epsffile{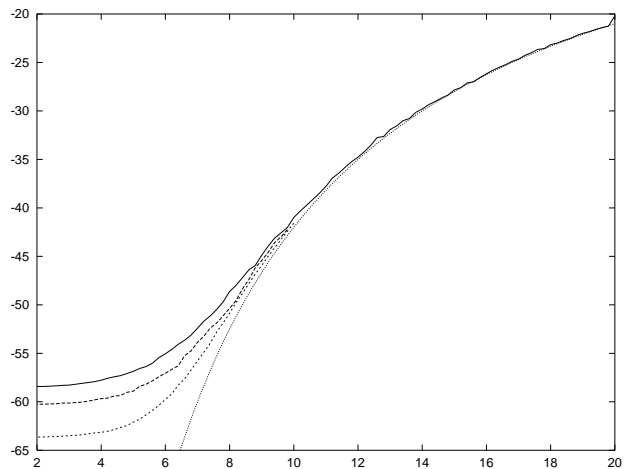} 
\end{center}
   \caption[0]{\protect\label{eneT_med}  
The energy for $N=4$ and three different values of the cooling 
rate ($2^{-5}$, $2^{-7}$, $2^{-12}$),
together with the first term of the high temperature 
expansion. The curves stay close to the first term 
of the high temperature expansion
until very close to where they fall off equilibrium.   
 } \end{figure}

Let us start with the discussion of cooling experiments.
In figure \ref{energie} and \ref{eneT_med} we show the energy as a function 
of the 
temperature for different cooling rates for $N=3$ and 4 respectively.
 The cooling rate $\kappa$ is equal to the inverse of the number of Monte Carlo
sweeps done at each temperature.
We recognize in both cases the typical curves of systems undergoing 
a glass transition, and remaining frozen below a cooling rate dependent 
freezing temperature. In the figure with $N=4$ we have plotted 
for comparison purposes the line corresponding to the first term
of the high temperature expansion. We see that until quite near
to the freezing the energy of the system remains close to that line. 
This is reminiscent to what happens in mean-field, 
where the line is followed up to the transition point. 

\begin{figure}
 \epsfxsize=250pt
\begin{center}
\epsffile{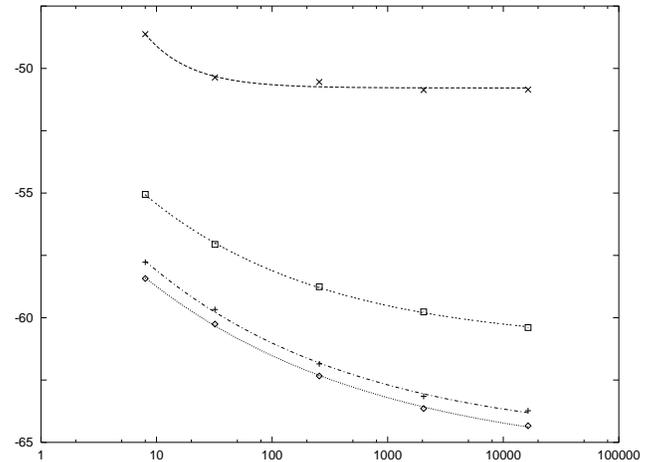} 
\end{center}
   \caption[0]{\protect\label{E-k}  
The energy for $N=4$ as a function of the inverse cooling rate. 
From top to bottom $T=8,6,4,2$. The full lines are the power law fits 
discussed in the text. 

 } \end{figure}

In figure \ref{E-k} we study the 
dependence of  the energy on the cooling rate for the 
$N=4$ model for $T=2$ and $T=4$ as a function of the inverse cooling rate.
The data are compatible with a power law relaxation of the 
kind $E(\kappa)=E_\infty +A \kappa^u$ with a temperature dependent 
exponent $u$. 
For instance a fit of the data for $T=2,4$ gives
 $E(\kappa)|_{T=2}=-65.8 + 11.6 \times \kappa^{0.21}$ and 
$E(\kappa)|_{T=4}=-65.1+11.9\times \kappa^{0.23}$. 
This dependence contrasts with the much slower logarithmic
dependence observed in real glasses and is the first sign of criticality 
in the system. 

\begin{figure}
 \epsfxsize=250pt
\begin{center}
\epsffile{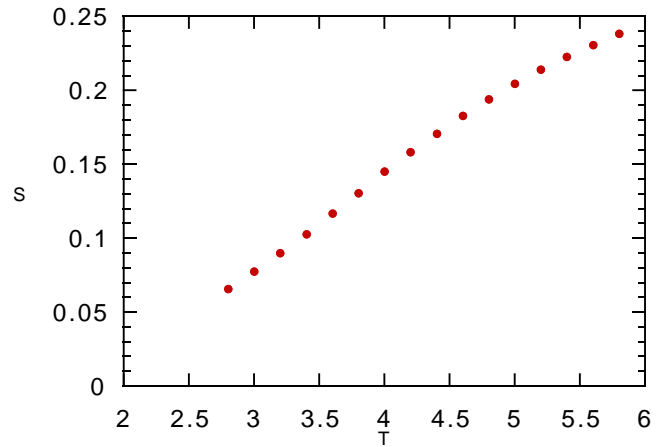} 
\end{center}
   \caption[0]{\protect\label{entropia_17}  
The entropy as a function of the temperature for $N=3$.
 } \end{figure}

In the equilibrium regime (which is reached for sufficient long simulations) 
we can obtain free-energy and entropy
integrating the data of the internal energy and taking into account that
at infinite temperature the entropy per spin is $S(T=\infty)=\log 2$.
The free-energy is reconstructed as 
\be
F=\log 2-T\int_0^{1/T} E(\b) \d \b,
\label{F}
\ee
 and from it the entropy, that 
we plot  in figure \ref{entropia_17}.
We observe that $S(T)$ behaves linearly in a wide range of temperatures
suggesting the validity of the Gibbs-Di Marzio transition mechanism for 
this model. As we will see the model with $N=3$ seems
to have a transition around 
$T=2.6$  while the linear extrapolation of the entropy vanishes only 
at $T\approx 1.8$. However it is not the total entropy, but the 
``configurational entropy'' (associated to the number of 
possible metastable states) that should vanish at the transition. 

\begin{figure}
 \epsfxsize=250pt
\begin{center}
\epsffile{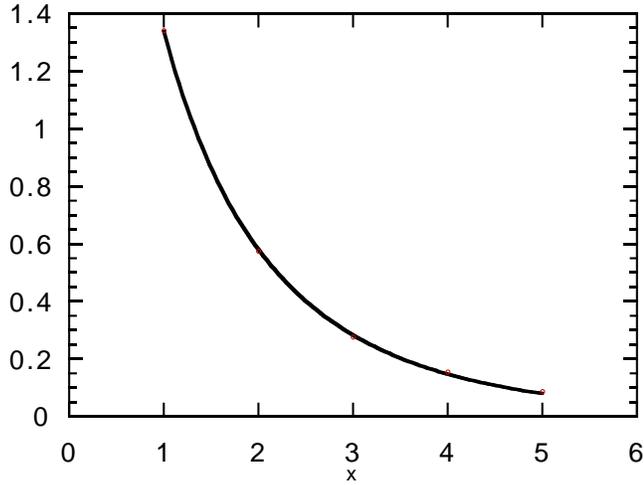} 
\end{center}
   \caption[0]{\protect\label{cor-spazioT.3}  
Overlap correlation function as a function of distance 
for $N=3$, $T=3$, together with the fit of the form 
$c(x)={A\over x+1} \exp(-x/\xi)$. The value of $\xi$ from the fit
is $\xi=2.5$. 
 } \end{figure}

\begin{figure}
 \epsfxsize=250pt
\begin{center}
\epsffile{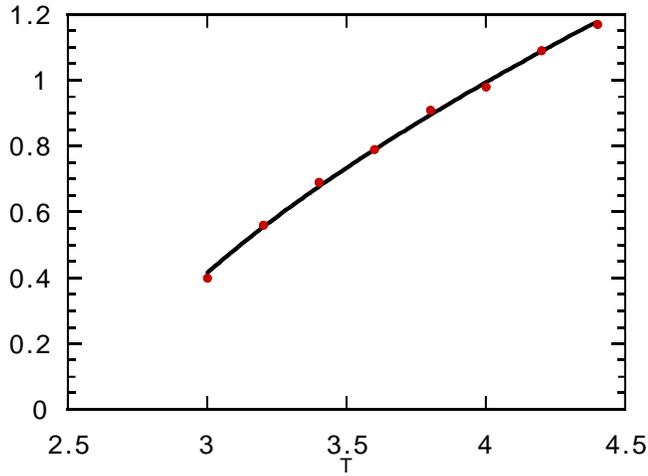} 
\end{center}
   \caption[0]{\protect\label{mass_2.63}  
Inverse of the correlation length versus temperature for $N=3$ and fit
$\xi^{-1}=(T-T_c)^\nu$, $\nu=0.71$, $T_c=2.62$.
 } \end{figure}

\begin{figure}
 \epsfxsize=250pt
\begin{center}
\epsffile{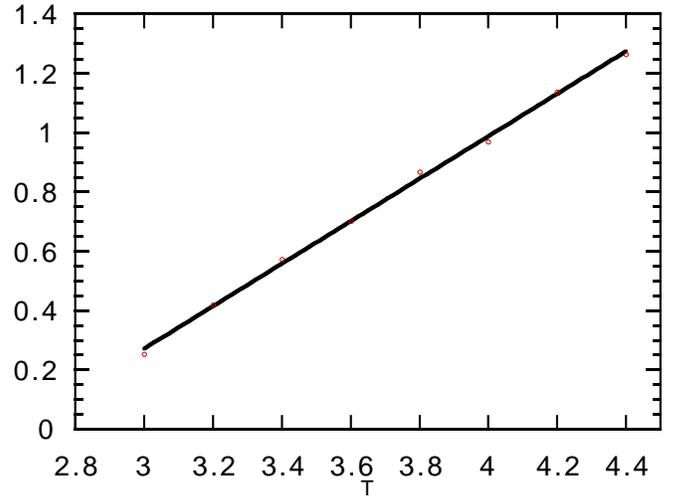} 
\end{center}
   \caption[0]{\protect\label{csi3over2f2}  
Correlation length $\xi$ to the power -2/3 versus temperature for $N=3$.
 The curve is a
linear fit.
 } \end{figure}

We now turn to the more difficult question of the identification
and  characterization of 
the phase transition in the model.
We have studied that issue  limiting ourselves  to the
case $N=3$. The quantity over which we have concentrated 
is the overlap correlation 
length. We simulated two identical replicas in parallel ($\sigma$ and $\tau$), 
and after thermalization we measured  the overlap correlation function:
\be 
G(x)={1\over N^2 V}
\sum_{i=1}^V\sum_{\a=1}^N\sum_{\b=1}^N \sigma_i^\a\sigma_{i+x}^\a
\tau_i^\b\tau_{i+x}^\b. 
\ee
The spin glass susceptibility is defined as 
\be 
\chi_{SG}=\int \d^3 x \; G(x).
\ee
The data for the function $G(x)$
are shown in figure \ref{cor-spazioT.3} for $T=3$, 
together with 
the the best fit of the form $G(x)=A/(x+1)^{1+\eta}\exp(-x/\xi)$
with $\eta=0$. We have done similar fits at different temperatures 
and in this way we have extracted the value of the  correlation length.
If we plot this correlation length as a function of the temperature 
(fig. \ref{mass_2.63}) we see that the data are best
fitted by the power form
$\xi\approx (T-T_c)^\nu$, with $T_c=2.62$ and $\nu=0.71$. However
from figure \ref{csi3over2f2} we see that the data are also well compatible 
with $\nu=2/3$ as one could expect from scaling arguments (see below).

Differently from mean field there is 
a correlation length growing in the system,  that suggests a
second order phase transition. 

\begin{figure}
 \epsfxsize=250pt
\begin{center}
\epsffile{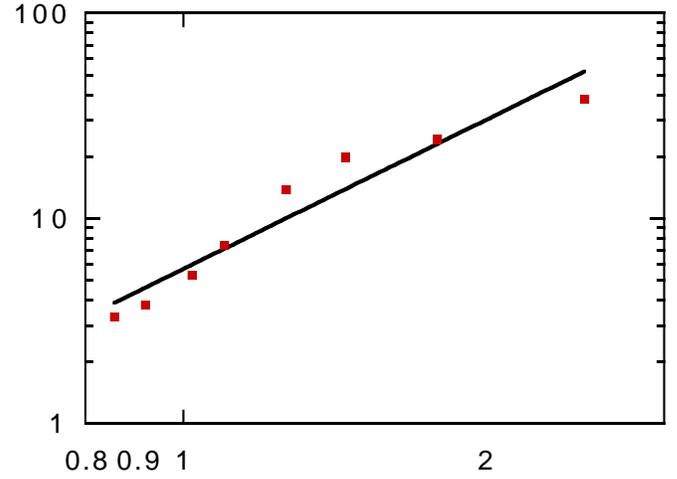} 
\end{center}
   \caption[0]{\protect\label{chiversuscspoweri2.4}  
Spin glass susceptibility $\chi_{SG}$ versus $\xi$ for $N=3$ 
and the power best fit which gives $\chi\sim \xi^{2.4}$
 } \end{figure}

Similar conclusions also come from the analysis of the spin glass 
susceptibility, which grows of about a factor ten in the range 
$3<T<4.4$. The data, shown in figure \ref{chiversuscspoweri2.4} are
roughly compatible with a power law divergence
of the susceptibility as $\sim \xi^{2.4}$, which using the 
value $2/3$ for $\nu$ corresponds to $\chi_{SG}\approx (T-T_c)^\gamma$
with $\gamma=1.6$. These values for the critical exponents $\gamma$ and
$\nu$ are definitely different from those of the Edwards Anderson spin glass
models: they are a factor 2-3 times smaller \cite{MPR}.

\begin{figure}
 \epsfxsize=250pt
\begin{center}
\epsffile{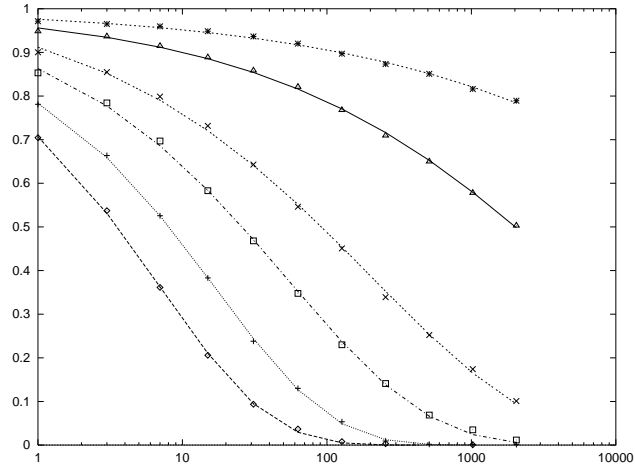} 
\end{center}
   \caption[0]{\protect\label{c-t}  
Time dependent auto-correlation function at equilibrium  in the $N=4$ model 
for, from top to bottom, $T=5,6,7,8,9,10$ the lines are strctched 
exponential fits. 
 } \end{figure}

\begin{figure}
 \epsfxsize=250pt
\begin{center}
\epsffile{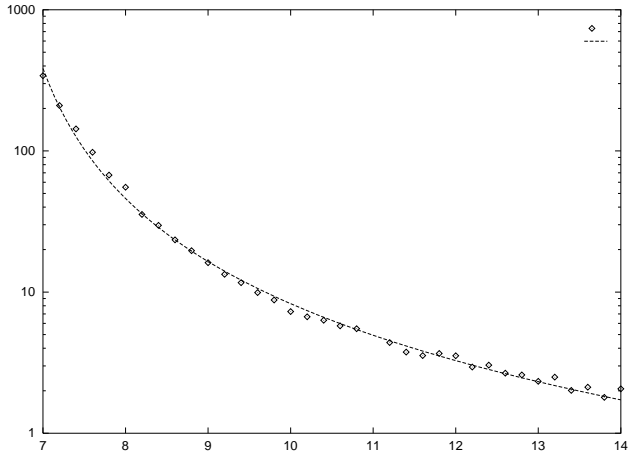} 
\end{center}~\caption[0]{\protect\label{tau2}  
Relaxation time extracted from the autocorrelation function 
as a function of the temperature for $N=4$.
The line is a power law fit $\tau=117.8\times (T-6.4)^2.1$.
 } \end{figure}

\begin{figure}
 \epsfxsize=250pt
\begin{center}
\epsffile{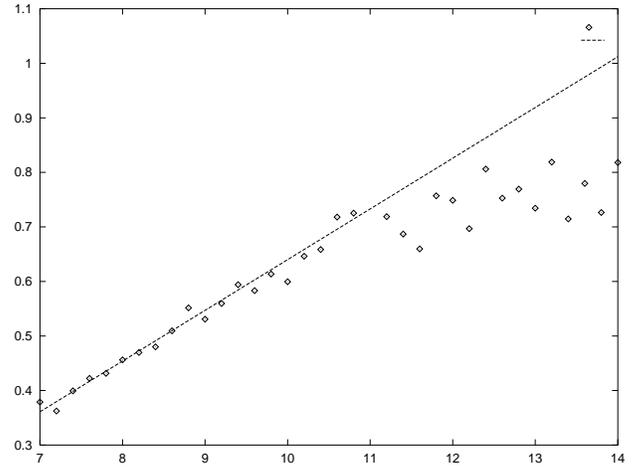} 
\end{center}~\caption[0]{\protect\label{beta}  
Stretching exponent versus temperature for  $N=4$.
The line is a linear fit of the region $T<10$: 
$\beta=T\times 0.093-0.29$.
 } \end{figure}

The last aspect that we have studied is the equilibrium relaxation,
and the relation of the relaxation time with the correlation length.
 In figure \ref{c-t}
we show the time autocorrelation function 
\be C(t) ={1\over NV}\sum_{i,\a} \sigma_i^\a(t)\sigma_i^\a(0) \ee
for various temperature for $N=4$. 
We see that a form $C(t)=A \exp(-(t/\tau)^\b)$
fits excellently the data. We extract from that the relaxation time
$\tau$  (figure \ref{tau2}) and 
the exponent $\beta$ that we plot in figure \ref{beta}.  

\begin{figure}
 \epsfxsize=250pt
\begin{center}
\epsffile{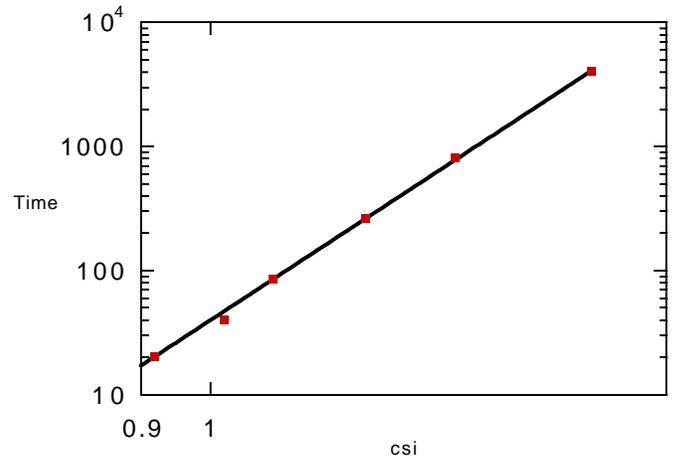} 
\end{center}
   \caption[0]{\protect\label{timeandcsi.power8}  
Relaxation time versus $\xi$ for $N=3$. The line is a power law fit 
with power $z=8$.} \end{figure}

We also done the same analysis for $N=3$ with similar results. 
It is interesting to plot $\tau$ {\it versus} $\xi$ which 
shows the compatibility of our data with the relation $\tau\sim
\xi^{z}$ (see figure \ref{timeandcsi.power8}). 
This scaling form is at variance with the results of \cite{frarit} for
the other short range p-spin model we mentioned in the introduction, 
and experiments in structural glasses. The high value of $z$ we find. 
i.e. $z=8$ 
(which by coincidence is quite similar to the value in the Edwards Anderson
spin glass model) implies a violent divergence of the correlation time near 
the transition and it is responsible of the large correlation time needed to
reach thermalization. 

The difference of behaviour with respect to the  mean-field 
can be rationalized with the  argument presented in the next section,
 which also implies 
$\nu=2/3$. 

\section{A possible theoretical interpretation}

The difference among the short range model and the homologous infinite range
 model (which can be solved in the mean field approximation) 
are quite striking.
No precursor signs of the transition are present in the infinite range model
and the spin glass susceptibility remains finite up to the transition point.
Here we will present and argument which suggests
that in short range models with quenched local disorder
the spin glass susceptibility is divergent as in second order 
phase transitions.

The precise  reasons for this discrepancy are not clear to us. The simplest
scenario  we have considered is the following. The disorder induces 
fluctuations in the local transition temperature. For a given region of
radius $R$ 
centered around the point $x$ we can define and effective critical temperature
$T_R(x)$. It is natural to assume that the $x$-dependent fluctuation 
of $T_R(x)$  
is a quantity  of order
$R^{-D/2}$, i.e.
\be
T_R(x)= T_c+\delta T_R(x)
\ee
with $\langle \delta T_R(x)^2 \rangle \propto R^D$. At a given temperature $T$
the
regions of size $R$ such that $T_R(x)>T$ are strongly correlated. 
The typical radius of these region increases as $(T-T_c)^{2/D}$
suggesting therefore that $\nu=2/D$. This value of $\nu$ is peculiar for random
systems. Indeed there are general arguments that show that for second
order phase
transitions disorder is relevant if $\nu \ge 2/D$ \cite{HARRIS}.

We also notice that the same value of $\nu$ can be obtained if we assume
that the specific heat has a discontinuity at the phase transition as predicted
by the mean field analysis. In other words 
we suppose that the specific heat exponent $\alpha$ is equal to zero. The usual
scaling law $\alpha=2-D \nu$ implies the result $\nu=2/D$. This  argument
is suggestive, but the coincidence its prediction with 
the value of $\nu$ we find may be fortuitous of the dimension three.
 An investigation of the model in 
higher dimensions will give some information on the validity on this conjecture
(numerical simulations in four dimensions for the $p=3$ model \cite{SEI}
suggest that in
this case $\nu$ is around .5).

\section{Conclusions}

In this paper we have studied by Monte Carlo a finite dimensional version 
of the p-spin model. We find a scenario for freezing that mixes 
typical features of structural glasses, like strong cooling rate dependence 
of the low temperature energy, with features of second order phase transitions
with power law growing of the correlation length and critical 
dynamics. This is a genuine finite dimensional effect due to the quenched
disorder which can be understood qualitatively with the argument we have 
given. A full theoretical understanding should come from the inclusion 
of non-perturbative effects in the theory. 
The simulations of disordered finite dimensional analogous of 
systems with ``one step replica breaking'' is just at the 
beginning and much progress can be expected in the future.

\end{document}